\documentclass[11pt]{article}
\usepackage{fullpage}
\usepackage{amsmath}
\usepackage{amssymb}
\usepackage{amsfonts}
\usepackage[all,poly]{xy}
\usepackage[dvips]{color}

\newcounter{algoctr}

\newif\ifnotesw\noteswtrue
   {\ifnotesw\marginpar[\hfill\(\top\)]{\(\top\)}\fi}%
   {\ifnotesw\marginpar[\hfill\(\bot\)]{\(\bot\)}\fi}

\newcommand{\mnote}[1]%
    {\ifnotesw\marginpar%
        [{\scriptsize\begin{minipage}[t]{\marginparwidth}
        \raggedleft#1%
                        \end{minipage}}]%
        {\scriptsize\begin{minipage}[t]{\marginparwidth}
        \raggedright#1%
                        \end{minipage}}%
    \fi}

\newcommand{\ignore}[1]{}

\newcommand{\etal}{{\it et al. }}

\newsavebox{\given}
\savebox{\given}[1em]{\rule[-1.5ex]{.2mm}{4ex}}

\newcommand{\bnum}{\begin{equation}}
\newcommand{\enum}{\end{equation}}

\newtheorem{theorem}{Theorem}
\newtheorem{corollary}[theorem]{Corollary}
\newtheorem{lemma}[theorem]{Lemma}

\newtheorem{fact}[theorem]{Fact}

\newcommand{\blackslug}{\rule{7pt}{7pt}}

\newcommand{\iverson}[1]{\lbrack\!\lbrack #1 \rbrack\!\rbrack}

\newcommand{\real}{\ifmmode {\rm R} \else ${\rm R}$ \fi}
\newcommand{\nat}{\ifmmode {\rm N} \else ${\rm N}$  \fi}
\newcommand{\tot}{\ifmmode {\cal T} \else ${\cal T}$ \fi}
\newcommand{\sigstar}{\ifmmode \Sigma^{\ast} \else $\Sigma^{\ast}$ \fi}
\renewcommand{\star}{\ast}

\newcommand{\inn}{\ifmmode \in \else $\in$ \fi}
\renewcommand{\phi}{\ifmmode \varphi \else $\varphi$ \fi}
\renewcommand{\le}{\ifmmode \leq \else $\leq$ \fi}
\renewcommand{\ge}{\ifmmode \geq \else $\geq$ \fi}
\renewcommand{\ne}{\ifmmode \neq \else $\neq$ \fi}
\newcommand{\lt}{\ifmmode < \else $<$ \fi}
\newcommand{\gt}{\ifmmode > \else $>$ \fi}
\newcommand{\eq}{\ifmmode = \else $=$ \fi}
\newcommand{\half}{\ifmmode \frac{1}{2} \else $\frac{1}{2}$ \fi}
\newcommand{\oneovern}{\ifmmode \frac{1}{n} \else $\frac{1}{n}$ \fi}

\newcommand{\ra}{\ifmmode \rightarrow \else $\rightarrow$ \fi}
\newcommand{\qed}{\hfill{\setlength{\fboxsep}{0pt}
\framebox[7pt]{\rule{0pt}{7pt}}}}

\renewcommand{\notin}{\ifmmode \not\in \else $\not\in$ \fi}
\newlength{\thislabel}
\newcommand{\labsize}[1]{\settowidth{\thislabel}{#1}}

\newcommand{\prf}{\par\noindent{\sl Proof } \hspace{.01 in}}
\newcommand{\zo}{\{0,1\}}

\newcommand{\lip}{\langle}
\newcommand{\rip}{\rangle}

\def\Complex{\mathbb C}
\def\Int{\mathbb Z}

\newcommand{\GG}{\mathcal{G}}
\newcommand{\XX}{\mathcal{X}}

\newcommand{\bra}[1]{\lip #1 |}
\newcommand{\ket}[1]{| #1 \rip}
\newcommand{\braket}[2]{\lip #1 | #2 \rip}

\newcommand{\dt}{\mathsf{d}t}

\title{
Non-uniform mixing of quantum walk on cycles
} 
\author{
{William Adamczak}\\{\em SUNY Albany} 
\and {Kevin Andrew}\\{\em Harvey Mudd College} 
\and {Leon Bergen}\\{\em Swarthmore College}
\and {Dillon Ethier}\\{\em Clarkson University}
\and {Peter Hernberg}\\{\em SUNY Potsdam} 
\and {Jennifer Lin}\\{\em Princeton University} 
\and {Christino Tamon}\footnote{Contact author: tino@clarkson.edu}\\{\em Clarkson University}
}
\date{\today}

\begin{document}
\bibliographystyle{plain}
\maketitle

\begin{abstract}
A classical lazy random walk on cycles is known to mix to the uniform distribution. In contrast,
we show that a continuous-time quantum walk on cycles exhibit strong non-uniform mixing properties.
Our results include the following:
\begin{itemize}
\item The {\em instantaneous} distribution of a quantum walk on most even-length cycles is never uniform. 
	More specifically, we prove that a quantum walk on a cycle $C_{n}$ is not instantaneous uniform mixing, 
	whenever $n$ satisfies either: (a) $n = 2^{u}$, for $u \ge 3$; 
	or (b) $n = 2^{u}q$, for $u \ge 1$ and $q \equiv 3\pmod{4}$. 
\item The {\em average} distribution of a quantum walk on any Abelian circulant graph is never uniform.
	As a corollary, the average distribution of a quantum walk on any standard circulant graph, 
	such as the cycles, complete graphs, and even hypercubes, is never uniform. 
	Nevertheless, we show that the average distribution of a quantum walk on the cycle $C_{n}$ 
	is $O(1/n)$-uniform.
\end{itemize}
\end{abstract}


\section{Introduction}

Quantum walk on graphs is a non-trivial and interesting generalization of classical random walk
on graphs. A mathematical theory of both had proved relevant to physics, computer science, and
more recently, to quantum information. An excellent survey of quantum walk on graphs is given
by Kendon \cite{k06}. In this work, we will focus on continuous-time unitary quantum walk on finite 
graphs. Our goal is to show strong non-uniform mixing properties of a continuous-time quantum
walk on cycles and circulant graphs, which demonstrates a distinct behavior from a classical
lazy random walk on the same graphs.

A continuous-time quantum walk on a graph $G=(V,E)$ is defined using Schr\"{o}dinger's equation
by treating the adjacency matrix of $G$ as the Hamiltonian of the quantum system. This treatment 
is standard in the physics literature (for example, see \cite{feynman}), where $G$ is commonly an 
infinite low-dimensional lattice. This corresponds to a quantum analogue of the important
investigations of classical random walks on $\mathbb{Z}^{d}$, for $d \ge 1$, by Polya and
others (see \cite{doyle-snell}).

Yet, the case when $G$ is a finite graph has only been analyzed recently due to its potential
applications in developing efficient quantum algorithms (see \cite{fg98,ccdfgs03}). An interesting 
mixing property of a continuous-time quantum walk on the hypercube graphs was observed by Moore and 
Russell \cite{mr02}.
They showed that a continuous-time quantum walk on the hypercube is instantaneous uniform mixing;
that is, there are times when the probability distribution of the quantum walk, when measured, equals 
exactly the uniform distribution on the vertices of the hypercube. Although a classical random walk
on the hypercube also mix to uniform, a quantum walk hits the uniform distribution asymptotically
faster.

Subsequent works showed that several other natural family of graphs do not share this uniform mixing 
property with the hypercube. For example, the complete graphs \cite{abtw03} and the Cayley graphs of
the symmetric group \cite{gw03} are known to be {\em not} instantaneous uniform mixing. 
But, there is a very natural class of graphs whose status remains open: the cycle graphs. 
Quantum walk on cycles had been studied in the discrete-time setting \cite{aakv01,bgklw03}. 
It is also known that the evolution of continuous-time quantum walk on cycles can be expressed as a 
summation involving Bessel functions (see \cite{feynman,bbt04}). 
Still, it is unknown if a continuous-time quantum walk on cycles has the uniform mixing property. 

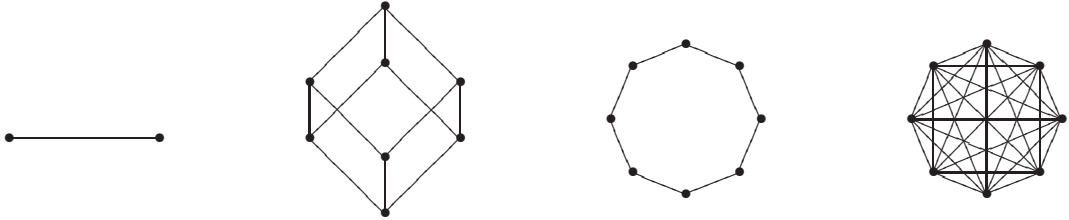
\begin{figure}[t]
\[\begin{xy} /r10mm/:
 ,{\xypolygon2"S"{~={0}\dir{*}}}
 ,+/r40mm/
 ,{\xypolygon4"M"{~={0}\dir{*}}}
 ,+(0.0,0.75),{\xypolygon4"N"{~={0}\dir{*}}}
 ,"M1";"N1"**@{-},"M2";"N2"**@{-},"M3";"N3"**@{-},"M4";"N4"**@{-}
 ,+/r40mm/
 ,+(0.0,0.5),{\xypolygon8"C"{~={0}\dir{*}}}
 ,+/r40mm/
 ,{\xypolygon8"K"{~={0}\dir{*}}}
 ,"K1";"K3"**@{-},"K2";"K4"**@{-},"K3";"K5"**@{-},"K4";"K6"**@{-},"K5";"K7"**@{-},"K6";"K8"**@{-},"K7";"K1"**@{-},"K8";"K2"**@{-}
 ,"K1";"K4"**@{-},"K2";"K5"**@{-},"K3";"K6"**@{-},"K4";"K7"**@{-},"K5";"K8"**@{-},"K6";"K1"**@{-},"K7";"K2"**@{-},"K8";"K3"**@{-}
 ,"K1";"K5"**@{-},"K2";"K6"**@{-},"K3";"K7"**@{-},"K4";"K8"**@{-}
\end{xy}\]
\caption{\small {\em Some Abelian circulants and their quantum mixing properties}.
From left to right: 
(a) the smallest circulant $K_{2}$; 
{instantaneous and average uniform}.
(b) the $3$-cube or $\mathbb{Z}_{2}^{3}$-circulant;
{instantaneous but not average uniform} (Moore and Russell \cite{mr02}).
(c) the cycle $C_{8}$ or sparse $\Int_{8}$-circulant; 
{not instantaneous uniform, but average $(1/n)$-uniform} (this work).
(d) the complete graph $K_{8}$ or dense $\Int_{8}$-circulant; 
{neither instantaneous nor average $(1/n)$-uniform} (Ahmadi \etal \cite{abtw03}).
}
\label{figure:abelian-circulant}
\vspace{.1in}
\hrule
\end{figure}

In this work, we show that a continuous-time quantum walk on cycles exhibit strong non-uniform mixing property.
First, we prove that on most even-length cycles the quantum walk is not instantaneous uniform mixing. 
The theorem applies to cycles whose lengths $n$ are either a power of two, say, $2^{u}$, where $u \ge 3$, 
or a product of a power of two and an odd number congruent to $3$ modulo $4$, that is, $n = 2^{u}q$, where
$u \ge 1$ and $q \equiv 3 \pmod{4}$. 
Our proof exploits spectral symmetries of even-length cycles coupled with some number-theoretic arguments.
In a sense, our arguments only yield non-uniform mixing for {\em half} of the even lengths; 
a separate argument seems needed for the case when $q \equiv 1\pmod{4}$.

Second, we consider average mixing of quantum walk on cycles. The notion of average mixing is a natural quantum 
generalization of stationary or limiting distributions in classical random walks on graphs. 
We prove a very general theorem stating that a continuous-time quantum walk on any Abelian group-theoretic 
circulant (defined in \cite{d90}) is not average uniform mixing. Since the class of Abelian circulants include
natural family of graphs such as the cycles, complete graphs, and even hypercubes, as a corollary, we obtain
the same non-uniform average mixing property for a continuous-time quantum walk on cycles. 
But, we also show that, in a quantum walk on cycles, the average distribution is $(1/n)$-close to uniform
(in total variation distance). This property is not true for the family of complete graphs.

Our work exploits heavily the circulant structure and spectral properties of the underlying graphs. 
A more complete treatment of circulants and their beautiful theory is given by Davis \cite{davis} and
Diaconis \cite{diaconis}, while a different aspect of quantum walk on circulant graphs is 
described by Saxena \etal \cite{sss07}.


\section{Preliminaries}

For a logical statement $S$, let $\iverson{S}$ denote the characteristic function of 
$S$ which evaluates to $1$ if $S$ is true, and to $0$ if it is false.

We consider only graphs $G=(V,E)$ that are simple, undirected, and connected. 
Let $A_{G}$ be the adjacency matrix of $G$, where $A_{G}[j,k] = \iverson{(j,k) \in E}$. 
A graph $G$ is {\em circulant} if its adjacency matrix $A_{G}$ is circulant. 
A circulant matrix $A$ is specified by its first row, say $[a_{0}, a_{1}, \ldots, a_{n-1}]$, and 
is defined as $A[j,k] = a_{k-j \pmod{n}}$, where $j,k \in \Int_{n}$:
\begin{equation}
A = \begin{bmatrix}
	a_{0} & a_{1} & \ldots & a_{n-1} \\
	a_{n-1} & a_{0} & \ldots & a_{n-2} \\
	\vdots & \vdots & \ldots & \vdots \\
	a_{1} & a_{2} & \ldots & a_{0}
	\end{bmatrix}
\end{equation}
Here $\Int_{n}$ denotes the group of integers $\{0,\ldots,n-1\}$ under addition modulo $n$.
Note that $a_{0} = 0$, since our graphs are simple, and $a_{j} = a_{n-j}$, since our graphs are undirected.
Most known families of circulant graphs include the complete graphs and cycles (see Figure \ref{figure:abelian-circulant}).
\ignore{
Alternatively, a circulant graph $G = (V,E)$ can be specified by a subset $S \subseteq \Int_{n}$, where $(j,k) \in E$
if $k-j \in S$. In this case, we write $G = \crc{S}$. We will assume that $S$ is closed under taking inverses,
namely, if $d \in S$, then $-d \in S$.
}

All circulant graphs $G$ are diagonalizable by the Fourier matrix $F$ whose columns $\ket{F_{k}}$
are defined as $\braket{j}{F_{k}} = \omega_{n}^{jk}/\sqrt{n}$, where $\omega_{n} = \exp(2\pi i/n)$:
\begin{equation}
F = \frac{1}{\sqrt{n}}
	\begin{bmatrix}
	1 & 1 & 1 & \ldots & 1 \\
	1 & \omega_{n} & \omega_{n}^{2} & \ldots & \omega_{n}^{n-1} \\
	1 & \omega_{n}^{2} & \omega_{n}^{4} & \ldots & \omega_{n}^{2(n-1)} \\
	\vdots & \vdots & \vdots & \ldots & \vdots \\
	1 & \omega_{n}^{n-1} & \omega_{n}^{2(n-1)} & \ldots & \omega_{n}^{(n-1)^{2}} 
	\end{bmatrix}
\end{equation}
In fact, we have $FAF^{-1} = \sqrt{n} \cdot diag(FA_{0})$, for any circulant $A$,
where $A_{0}=A\ket{0}$ is the first column of $A$ (see \cite{davis,biggs}). 
This shows that the eigenvalues of $A$ are given by
\begin{equation} \label{eqn:circulant-eigenvalue}
\lambda_{j} = \sum_{k=0}^{n-1} a_{k} \ \omega_{n}^{-jk}.
\end{equation}

A {\em continuous-time quantum walk} on a graph $G=(V,E)$ is defined using the Schr\"{o}dinger equation 
with the real symmetric matrix $A_{G}$ as the Hamiltonian (see Farhi and Gutmann \cite{fg98}). A classical
random walk on a $d$-regular graph $G$ is sensitive to the choice of the stochastic transition matrix: 
either $\frac{1}{d} A_{G}$, for the simple walk, or $\frac{1}{2}I + \frac{1}{2d}A_{G}$, for the lazy walk. 
In our quantum walk, this choice is irrelevant since $I$ and $A_{G}$ commute, which implies that
$e^{-it(\frac{1}{2}I + \frac{1}{2d}A_{G})} = e^{-it/2}e^{-i(t/2d)A_{G}}$. The first term $e^{-it/2}$ is 
an irrelevant phase factor, while the second term involves a time shift $t/2d$ that may also be ignored.
Thus, we may assume that our stochastic transition matrix is simply $A_{G}$.
If $\ket{\psi(t)} \in \Complex^{|V|}$ is a time-dependent amplitude vector on the vertices of $G$, 
then the evolution of the quantum walk is given by
\begin{equation}
\ket{\psi(t)} = e^{-it A_{G}} \ket{\psi(0)},	
\end{equation}
where $i = \sqrt{-1}$ and $\ket{\psi(0)}$ is the initial amplitude vector. 
The amplitude of the quantum walk of vertex $j$ at time $t$ is given by $\braket{j}{\psi(t)}$.
The {\em instantaneous} probability of vertex $j$ at time $t$ is $p_{j}(t) = |\braket{j}{\psi(t)}|^{2}$. 
Let $P_{t} = \langle p_{j}(t) : j \in V \rangle$ be the instantaneous probability distribution of the quantum walk.

The {\em average} probability of vertex $j$ is defined as 
\begin{equation}
\overline{p}_{j} = \lim_{T \rightarrow \infty} \frac{1}{T} \int_{0}^{T} p_{j}(t) \ \dt.
\end{equation}
The average probability distribution of the quantum walk will be denoted $\overline{P}$. 
This notion of average distribution (defined in \cite{aakv01} for discrete-time quantum walks)
is similar to the notion of a stationary distribution in classical random walks. 

Given two probability distributions $P,Q$ on a finite set $S$, the {\em total variation distance} 
between $P$ and $Q$ is defined as $||P - Q|| = \sum_{s \in S} |P(s) - Q(s)|$.
Let $U$ be the uniform distribution on the vertices $V$ of $G$.
For a given $\varepsilon \ge 0$, we say that $G$ is {\em instantaneous} $\varepsilon$-uniform {\em mixing} 
if there is a time $t$ so that the total variation distance between $P_{t}$ and $U$ is at most $\varepsilon$,
that is, $||P_{t} - U|| \le \varepsilon$.
We also say that $G$ is {\em average $\varepsilon$-uniform mixing} if the total variation distance between
$\overline{P}$ and $U$ is at most $\varepsilon$, that is, $||\overline{P} - U|| \le \varepsilon$.
Whenever $\varepsilon = 0$, we say that {\em exact} uniform mixing is achieved.

For discrete-time quantum walk, Aharonov \etal \cite{aakv01} showed that a graph with distinct eigenvalues 
is potentially average uniform mixing. A continuous-time adaptation of this result is as follows. 
Suppose that $G$ has eigenvalues $\lambda_{0} \ge \ldots \ge \lambda_{n-1}$ with corresponding orthonormal 
eigenvectors $\ket{z_{0}}, \ldots, \ket{z_{n-1}}$.
The average probability of vertex $\ell$ is
\begin{eqnarray}
\overline{P}(\ell) 
	& = &
	\lim_{T \rightarrow \infty} \frac{1}{T} \int_{0}^{T} |\bra{\ell} e^{-itH} \ket{\psi(0)}|^{2} \dt  \\
	& = &
	\sum_{j,k=0}^{n-1} \braket{z_{j}}{0}\braket{0}{z_{k}}\braket{\ell}{z_{j}}\braket{z_{k}}{\ell}
		\lim_{T \rightarrow \infty} \frac{1}{T} \int_{0}^{T} e^{-it(\lambda_{j}-\lambda_{k})} \dt.
\end{eqnarray}
Since $\lim_{T \rightarrow \infty} \frac{1}{T}\int_{0}^{T} e^{-it \Delta} \dt = \iverson{\Delta = 0}$,
this implies that
\begin{equation} \label{eqn:aakv}
\overline{P}(\ell) =
	\sum_{j,k=0}^{n-1} \braket{z_{j}}{0}\braket{0}{z_{k}}\braket{\ell}{z_{j}}\braket{z_{k}}{\ell}
		\iverson{\lambda_{j} = \lambda_{k}},
\end{equation}
Moreover, if all eigenvalues are distinct, then
$\overline{P}(\ell) = \sum_{j=0}^{n-1} |\braket{\ell}{z_{j}}|^{2} |\braket{z_{j}}{0}|^{2}$.
\ignore{
So, if all eigenvalues are distinct, then average uniform mixing is achieved, under mild conditions on 
the eigenvectors. In particular, there is hope if the graph is diagonalized by the Fourier or Hadamard matrix;
however, these graphs do not have distinct eigenvalues, as we observe later. Nevertheless, 
Aharonov \etal \cite{aakv01} proved that {\em discrete} quantum walks on odd cycles are average uniform mixing.
}


\section{Non-uniform instantaneous mixing of even-length cycles}

In this section, we show that a continuous-time quantum walk on most even-length cycles $C_{n}$ is not 
instantaneous uniform mixing.
Using Equation (\ref{eqn:circulant-eigenvalue}), the eigenvalues of a cycle $C_{n}$ are given by
\begin{equation}
\lambda_{k} = 2\cos(2\pi k/n), \ \ \ k = 0,\ldots,n-1.
\end{equation}
Note that $\lambda_{0} = 2$, $\lambda_{n-k} = \lambda_{k}$, for $1 \le k < n/2$, and $\lambda_{n/2} = -2$,
when $n$ is even.
Let $\ket{\psi_{n}(t)}$ describe a continuous-time quantum walk on $C_{n}$ starting at vertex $0$. 
If $A$ is the circulant adjacency matrix of $C_{n}$, then
\begin{equation}
\ket{\psi_{n}(t)}
	= e^{-iAt}\ket{0} = e^{-iAt} \sum_{k=0}^{n-1} \frac{1}{\sqrt{n}} \ket{F_{k}} 
	= \frac{1}{\sqrt{n}}\sum_{k=0}^{n-1} e^{-i\lambda_{k}t} \ket{F_{k}}.
\end{equation}
This shows that, for each $j=0,\ldots,n-1$, we have
\begin{equation} \label{eqn:circulant}
\braket{j}{\psi_{n}(t)} = \frac{1}{n}\sum_{0 \le k < n} e^{-i\lambda_{k}t} \omega_{n}^{jk},
\end{equation}

\begin{fact} \label{fact:amplitude}
Let $\ket{\psi_{n}(t)}$ describe a continuous-time quantum walk on $C_{n}$, where $n$ is even. Then, for any
$j=0,\ldots,n-1$, we have
\begin{equation}
\braket{j}{\psi_{n}(t)} 
	= \frac{1}{n}\left\{e^{-2it} + (-1)^{j}e^{2it}
		+ 2\sum_{1 \le k < n/2} e^{-i\lambda_{k}t}\cos(2\pi jk/n)\right\}.
\end{equation}
\end{fact}
\prf
Using the eigenvalue symmetry $\lambda_{k} = \lambda_{n-k}$, for $1 \le k < n/2$, combined with 
Equation (\ref{eqn:circulant}), yields the claim.
\ignore{
\begin{equation}
\braket{j}{\psi_{n}(t)}
	= \frac{1}{n} \left\{e^{-2it} + (-1)^{j} e^{2it} + 2\sum_{1 \le k < n/2} e^{-i\lambda_{k}t} \cos(2\pi jk/n) \right\}.
\end{equation}
}
\qed\\

The following lemma shows that some properties of a quantum walk on $C_{n}$ can be deduced from a quantum walk on $C_{m}$, 
if $m$ divides $n$. This reduction will be helful in analyzing a quantum walk on even-length cycles.

\begin{lemma} \label{lemma:decomposition}
Let $m,n > 0$ be integers so that $m | n$. Then, for each $0 \le a < m$ we have
\begin{equation}
\sum_{0 \le j < n} \iverson{j \equiv a \ (\mbox{mod } m)} \braket{j}{\psi_{n}(t)} = \braket{a}{\psi_{m}(t)}
\end{equation}
\end{lemma}
\prf
Using Equation (\ref{eqn:circulant}), after switching summations, we get:
\begin{equation}
	\sum_{0 \le j < n} \iverson{j \equiv a \ (\mbox{mod } m)} \braket{j}{\psi_{n}(t)}
	= 
	\frac{1}{n}\sum_{0 \le k < n} e^{-i\lambda_{k}t} 
		\sum_{0 \le j < n} \iverson{j \equiv a \ (\mbox{mod } m)} \times \omega_{n}^{jk}.
\end{equation}
Rewriting the inner index $j$ as $m\tilde{j} + a$, as $\tilde{j}$ vary in $0 \le \tilde{j} < n/m$, we get
(after renaming $\tilde{j}$ back to $j$):
\begin{equation}
	\frac{1}{n}\sum_{0 \le k < n} e^{-i\lambda_{k}t} 
		\sum_{0 \le j < n/m} \omega_{n}^{(mj+a)k} 
	= \frac{1}{n}\sum_{0 \le k < n} e^{-i\lambda_{k}t} \omega_{n}^{ak}
		\sum_{0 \le j < n/m} \omega_{n/m}^{jk}.
\end{equation}
Next, we note that $\sum_{j=0}^{n-1} \omega_{n}^{j} = n \times \iverson{k \equiv 0\pmod{n}}$. This yields:
\begin{equation}
	\frac{1}{m}\sum_{0 \le k < n} \iverson{k \equiv 0 \ (\mbox{mod } n/m)} \times e^{-i\lambda_{k}t} \omega_{n}^{ak}
	= \frac{1}{m}\sum_{0 \le k < m} e^{-i\lambda_{k}t} \omega_{m}^{ak},
\end{equation}
which equals $\braket{a}{\psi_{m}(t)}$.
\qed\\

A quantum walk on $C_{2}$, which is a multigraph on two vertices with two distinct edges connecting the vertices, is given by:
\begin{equation}
\ket{\psi_{t}} = \exp\left(-it\begin{bmatrix} 0 & 2 \\ 2 & 0 \end{bmatrix}\right) \ket{0}
	= \frac{1}{2} 
	\left\{e^{-2it} \begin{bmatrix} 1 \\ 1 \end{bmatrix} + e^{2it} \begin{bmatrix} 1 \\ -1 \end{bmatrix} \right\} 
	= \begin{bmatrix} \cos(2t) \\ -i\sin(2t) \end{bmatrix}.
\end{equation}
Thus, we have $P_{t} = \begin{bmatrix} \cos^{2}(2t) \ \sin^{2}(2t) \end{bmatrix}^{T}$.
Applying the previous lemma to even-length cycles, a quantum walk on $C_{2n}$ behaves in a similar manner to a
quantum walk on a $2$-vertex cycle. More specifically, the sum of the amplitudes on the vertices with even 
(respectively, odd) indices in a quantum walk on $C_{2n}$ corresponds exactly to the amplitude of vertex $0$ 
(respectively, $1$) in a quantum walk on $C_{2}$.

\begin{corollary} \label{corollary:cos-sin}
Let $\ket{\psi_{n}(t)}$ describe a continuous-time quantum walk on $C_{n}$, where $n$ is even. Then,
\begin{equation}
\sum_{0 \le j < n/2} \braket{2j}{\psi_{n}(t)} = \cos(2t), \ \ \ \ \
\sum_{0 \le j < n/2} \braket{2j+1}{\psi_{n}(t)} = -i\sin(2t)
\end{equation}
\end{corollary}
\prf
Since $n$ is even, apply Lemma \ref{lemma:decomposition} with $m=2$.
\qed\\

A further eigenvalue symmetry on even-length cycles yields a useful simplification on the amplitude expression
given by Fact \ref{fact:amplitude}.

\begin{lemma} \label{lemma:quarter}
Let $\ket{\psi_{n}(t)}$ describe a continuous-time quantum walk on $C_{n}$, where $n$ is even.
Then,
\begin{equation}
\braket{j}{\psi_{n}(t)} = 
	\frac{1}{n} \left\{ \varepsilon^{(n)}_{j,0}(t) +
		2\sum_{1 \le k < n/4} \varepsilon^{(n)}_{j,k}(t) \cos(2\pi jk/n) \right\},
\end{equation}
where $\varepsilon^{(n)}_{j,k}(t) = e^{-i\lambda_{k}t} + (-1)^{j}e^{i\lambda_{k}t}$.
\end{lemma}
\prf
Using Fact \ref{fact:amplitude} and the eigenvalue symmetry $\lambda_{n/2-k} = -\lambda_{k}$, 
for $1 \le k < n/4$, we obtain
\begin{eqnarray}
\braket{j}{\psi_{n}(t)} 
	& = & \frac{1}{n} \left\{e^{-2it} + (-1)^{j}e^{2it} + 
		2\sum_{1 \le k < n/2} e^{-i\lambda_{k}t} \cos(2\pi jk/n) \right\} \\
	& = & \frac{1}{n} \left\{e^{-2it} + (-1)^{j}e^{2it} + 
		2\sum_{1 \le k < n/4} (e^{-i\lambda_{k}t} + (-1)^{j}e^{i\lambda_{k}t}) \cos(2\pi jk/n) \right\}
\end{eqnarray}
since $\cos(2\pi j(n/2-k)/n) = (-1)^{j}\cos(2\pi jk/n)$. 
\qed\\

Using the previous lemma, we may deduce that the amplitude values on the vertices in a quantum walk on an even-length
cycle are purely real or purely imaginary; moreover, this is completely determined by the parity of the vertex index.

\begin{corollary} \label{cor:real-imag}
Let $\ket{\psi_{n}(t)}$ describe a continuous-time quantum walk on $C_{n}$, where $n$ is even.
Then, $\braket{j}{\psi_{n}(t)}$ is a real number, if $j$ is even, and is an imaginary number, if $j$ is odd.
\end{corollary}
\prf
Using Lemma \ref{lemma:quarter}, we note that $e^{-i\lambda_{k}t} + (-1)^{j}e^{i\lambda_{k}t}$ equals 
$2\cos(\lambda_{k}t)$, whenever $j$ is even, and equals $-2i\sin(\lambda_{k}t)$, if $j$ is odd.
\qed\\ 

Next, we show a lemma which connects the sum of amplitudes on a pair of vertices in a quantum walk on
$C_{2n}$ to the amplitude on a single vertex in a quantum walk on $C_{n}$. This lemma will be useful in
deducing the type of amplitude expressions involved in a quantum walk on $C_{2^{u}}$, for some $u \ge 1$.

\begin{lemma} \label{lemma:two-one}
Let $\ket{\psi_{n}(t)}$ describe a continuous-time quantum walk on $C_{n}$, where $n$ is even. Then,
for all $0 \le j < n$, we have
\begin{equation}
\braket{j}{\psi_{2n}(t)} + \braket{n-j}{\psi_{2n}(t)} = \braket{j}{\psi_{n}(t)}.
\end{equation}
\end{lemma}
\prf
Note that $\cos(2\pi(n-j)k/(2n)) = (-1)^{k}\cos(2\pi jk/(2n))$. 
By Lemma \ref{lemma:quarter}, the sum $\braket{j}{\psi_{2n}(t)} + \braket{n-j}{\psi_{2n}(t)}$ equals
\begin{equation}
	\frac{1}{n} \left\{\varepsilon^{(2n)}_{j,0}(t) + 
		2\sum_{1 \le k < n/2} \iverson{k \mbox{ even}} \cos\left(\frac{2\pi jk}{2n}\right) 
		\varepsilon^{(2n)}_{j,k}(t) \right\}
\end{equation}
Since $\varepsilon^{(2n)}_{j,2k}(t) = \varepsilon^{(n)}_{j,k}(t)$, we get
\begin{equation}
	\frac{1}{n} \left\{\delta^{(n)}_{j,0} + 
		2\sum_{1 \le k < n/4} \cos(2\pi jk/n) \varepsilon^{(n)}_{j,k}(t) \right\}.
\end{equation}
Again by Lemma \ref{lemma:quarter}, the last expression equals $\braket{j}{\psi_{n}(t)}$.
\qed\\

Finally, we are ready to state and prove a theorem showing that a continuous-time quantum walk most even-length
cycles is {\em not} instantaneous exactly uniform mixing. The proof uses several observations stated in the
previous lemmas.

\begin{theorem}
The family of cycles $C_{n}$ is not instantaneous uniform mixing, where $n$ satisfies
either (a) $n = 2^{u}$, where $u \ge 3$; or (b) $n = 2^{u}q$, where $u \ge 1$ and $q \equiv 3\pmod{4}$.
\end{theorem}
\prf
Assume that there is a time $t$ for which $|\braket{j}{\psi(t)}|^{2} = 1/n$.
By Corollary \ref{cor:real-imag}, we have that
\begin{equation}
\braket{j}{\psi(t)} =
	\left\{\begin{array}{ll}
		\pm 1/\sqrt{n}	& \mbox{ if $j$ is even } \\
		\pm i/\sqrt{n}	& \mbox{ if $j$ is odd }
	\end{array}\right.
\end{equation}
Also, we have
\begin{equation}
\cos(2t) = \sum_{0 \le j < n/2} \braket{2j}{\psi(t)} 
	= \frac{n}{2} \frac{1}{\sqrt{n}} - \frac{2k}{\sqrt{n}}
	= \frac{(n-4k)}{2\sqrt{n}},
\end{equation}
where $k$ is the number of $j$'s for which $\braket{2j}{\psi(t)}$ is negative.
Similarly, we have
\begin{equation}
-i\sin(2t) = \frac{(n-4\ell)}{2\sqrt{n}},
\end{equation}
where $\ell$ is the number of $j$'s for which $\braket{2j+1}{\psi(t)}$ is negative.
Since $\cos^2(2t) + \sin^2(2t) = 1$, we obtain
\begin{equation} \label{eqn:reduction}
(n-4k)^{2} + (n-4\ell)^{2} = 4n.
\end{equation}
Let $a_{k} = n-4k$ and $a_{\ell} = n-4\ell$.
There are two cases to consider: one of $a_{k}$ or $a_{\ell}$ is zero, or both are non-zero. 

If one of them is zero, say $a_{k}=0$, then $a^{2}_{\ell} = 4n$. If $n = 2^{u}q$, where $u \ge 1$
and $q \equiv 3\pmod{4}$, we have a contradiction since $2^{u}q$ is not a square, if $u$ is odd
or $q \equiv 3\pmod{4}$.
Otherwise, $u$ is even and $q=1$,
and, by repeated applications of Lemma \ref{lemma:two-one}, we observe that
$\braket{0}{\psi_{8}(t)} = a/\sqrt{n} = a/2^{m}$,
for some integers $a,m \in \mathbb{Z}$. But, by Lemma \ref{lemma:quarter}, we have
\begin{equation}
\braket{0}{\psi_{8}(t)} = \frac{1}{4}\cos(2t) + \frac{1}{2}\cos(\sqrt{2}t).
\end{equation}
Since $\cos(t)=0$ or $\sin(t)=0$ in this case, we must have $t = k(\pi/2)$, for some $k \in \mathbb{Z}$.
But $\cos(k\pi/\sqrt{2})$ is not rational, for any integer $k$, since $(e^{i(\pi/2)\sqrt{2}})^{k}$ is 
transcendental, by the Gelfond-Schneider\footnote{The Gelfond-Schneider theorem states that
$\alpha^{\beta}$ is transcendental, if $\alpha$ and $\beta$ are algebraic numbers with
$\alpha \neq 0$ and $\alpha \neq 1$, and if $\beta$ is not a real rational number.}
theorem (see \cite{niven}).

Next, we consider the case when both $a_{k}$ and $a_{\ell}$ are non-zero. If both terms are odd, then
considering Equation (\ref{eqn:reduction}), the left-hand size satisfies $a_{k}^{2} + a_{\ell}^{2} \equiv 2\pmod{4}$ 
whereas the right-hand side satisfies $4n \equiv 0\pmod{4}$; this is a contradiction. Otherwise, if both terms are even, 
then a factor of $4$ can be removed from both sides of Equation (\ref{eqn:reduction}). Continuing this process, we
arrived at a case where either the right-hand side is $q \equiv 3\pmod{4}$ or both $a_{k}$ and $a_{\ell}$ are odd. 
In either case, we arrive at a contradiction modulo $4$.

This completes the proof of the theorem.
\qed\\


\section{Non-uniform average mixing on Abelian circulants}

Our main theorem in this section shows that the average distribution of a continuous-time quantum walk
on any cycle, except for $C_{2}$, is never uniform. In fact, we prove a much stronger theorem stating
that the average distribution of a continuous-time quantum walk on any $\GG$-circulant, for any Abelian
group $\GG$ (as defined in \cite{d90}) is never uniform, except for $C_{2}$.

Diaconis \cite{d90} described the following interesting group-theoretic generalization of circulants. 
Let $\GG$ be a group of order $n$ and let $f: \GG \rightarrow \Complex$ be a {\em class} function of $\GG$
(that is, it is constant on the conjugacy classes of $\GG$).
Consider the matrix $M_{\GG}^{f}$ defined on $\GG \times \GG$ as $M_{\GG}^{f}[s,t] = f(st^{-1})$.
Note that with $\GG$ being the cyclic group $\Int_{n}$ of order $n$, we recover the standard 
circulant graphs, whereas with $\GG = \Int_{2}^{n}$, we obtain the hypercube graphs. 

Let $\rho: \GG \rightarrow GL(n,\Complex)$ be a representation of $\GG$ with dimension $n$.
The Fourier transform of $f$ at a representation $\rho$ is defined as
\begin{equation}
\hat{f}(\rho) = \sum_{x \in \GG} f(x)\rho(x).
\end{equation}
\ignore{
If $g : \GG \rightarrow \Complex$, then the convolution of $f$ and $g$ is
\begin{equation}
(f \star g)(y) = \sum_{x \in \GG} f(yx^{-1})g(x).
\end{equation}
The Fourier transform of a convolution is the product of the Fourier transforms:
\begin{equation}
\widehat{f \star g}(\rho) = \hat{f}(\rho)\hat{g}(\rho).
\end{equation}
}
As usual, Fourier inversion reconstructs $f$ from its Fourier transform at all irreducible representations 
$\rho_{1},\ldots,\rho_{m}$ of $\GG$ with dimensions $d_{1},\ldots,d_{m}$, respectively:
\begin{equation}
f(x) = \frac{1}{|\GG|} \sum_{j=1}^{m} d_{j} Trace(\rho_{j}(x^{-1})\hat{f}(\rho_{j})).
\end{equation}
For each irreducible representation $\rho_{j}$, we define a $d_{j}^{2} \times d_{j}^{2}$ block matrix $D_{j}$
as $D_{j} = diag(\hat{f}(\rho_{j}))$. Next, let $D = diag(D_{1},\ldots,D_{h})$ be a $|\GG| \times |\GG|$ matrix,
since $|\GG| = \sum_{j=1}^{m} d_{j}^{2}$.
Also, define the vector $\psi_{j}$ of length $d_{j}^{2}$ as
$\psi_{j}(x) = \frac{\sqrt{d_{j}}}{|\GG|} \langle \rho_{j}(x)[s,t] : 1 \le s,t \le d_{j} \rangle$
and the vector $\psi(x) = \langle \psi_{j}(x) : 1 \le j \le m \rangle$ of length $|\GG|$. 
Finally, we define the matrix $\XX = [\psi(x_{1}) \ldots \psi(x_{n})]$, where
$x_{1},\ldots,x_{n}$ are the elements of $\GG$.

\begin{theorem} \label{thm:diaconis} (Diaconis \cite{d90}) 
If $f: \GG \rightarrow \Complex$ is a class function of a finite group $\GG$, 
then $M_{\GG}^{f}$ is {\em unitarily diagonalized} by $\XX$, that is,
$M_{\GG}^{f} = \XX^{\dagger} D \XX$,
where, for each $j = 1,\ldots,m$, we have $D_{j} = \lambda_{j} I_{d_{j}^{2}}$, 
$\chi_{j}(x) = Trace(\rho_{j}(x))$ is the character of $\rho_{j}$ at $x$,
and the eigenvalue is 
\begin{equation} \label{eqn:group_lambda}
\lambda_{j} = \frac{1}{d_{j}}\sum_{x \in \GG} f(x)\overline{\chi_{j}}(x).
\end{equation}
\end{theorem}

We are interested in applying Theorem \ref{thm:diaconis} for an Abelian group $\GG$, where
all of its group representations have dimension one. Our main result shows that the spectral
gap of $M_{\GG}^{f}$, for any Abelian group $\GG$, is zero. This shows that average uniform
mixing is impossible.

The Hadamard matrix $H_{n}$ (of Sylvester type) is defined recursively as:
\begin{equation} \label{eqn:hadamard}
H_{2} = \begin{bmatrix} 1 & 1 \\ 1 & -1 \end{bmatrix}, \ \ \mbox{ and } \ \
H_{n} = \begin{bmatrix} H_{n-1} & H_{n-1} \\ H_{n-1} & -H_{n-1} \end{bmatrix}, \ \ \ n > 2.
\end{equation}
We call graph $G$ a {\em Hadamard} circulant if it is diagonalized by some Hadamard matrix $H_{n}$.
Alternatively, these are $\GG$-circulant matrices for $\GG = \Int_{2}^{n}$.
\ignore{
The following result was a main result in \cite{mr02}.

\begin{theorem} \label{thm:moore-russell} (Moore, Russell \cite{mr02})
A continuous-time quantum walk on the $n$-cube is instantaneous exactly uniform for times
$t = k \frac{\pi}{4} n$, for odd positive integers $k$. Also, there is $\varepsilon > 0$ such that
no $\varepsilon$-average mixing exists.
\end{theorem}
}
Although Equation (\ref{eqn:aakv}) suggests that a graph with distinct eigenvalues diagonalized 
by Hadamard matrices might be average uniform mixing, the following lemma disproves this possibility.

\begin{lemma} \label{lemma:hadamard}
Let $G$ be a graph diagonalized by a Hadamard matrix $H_{n}$, for $n > 2$. 
Then $G$ has spectral gap zero.
\end{lemma}
\prf 
Consider the characters of $\Int_{2}^{n}$ defined for each $a \in \Int_{2}^{n}$ as 
$\chi_{a}(x) = \prod_{j=1}^{n} (1-2a_{j}x_{j})$. From Equation (\ref{eqn:group_lambda}), we get
$\lambda_{a} = \sum_{x \in \Int_{2}^{n}} f(x)\chi_{a}(x)$,
where $f: \Int_{2}^{n} \rightarrow \zo$ defines the first column of the adjacency matrix of $G$. 
Let $|f| = \{x \neq 0_{n} : f(x) = 1\}$. Assume that $|f| < 2^{n}-1$, otherwise we get the complete graph
which has only $2$ distinct eigenvalues.
If $|f|$ is even, then $\lambda_{a} \in \{0, \pm 2, \ldots, \pm |f|\}$. Since the eigenvalues can take at most 
$|f|+1 < 2^{n}$ values, by the pigeonhole principle, there exist two non-distinct eigenvalues.
If $|f|$ is odd, then $\lambda_{a} \in \{\pm 1, \pm 3, \ldots, \pm |f|\}$. Similarly, the eigenvalues range
on at most $|f| < 2^{n}-1$ values, and again there exist two non-distinct eigenvalues.
\qed

\begin{theorem} \label{thm:abelian_circulant}
For any Abelian group $\GG$, no $\GG$-circulant, except for $C_{2}$, is average uniform mixing.
\end{theorem}
\prf Let $\GG = \Int_{n_{1}} \times \ldots \times \Int_{n_{k}}$ be an Abelian group.
If all elements of $\GG$ have order $2$ (except for the identity), we appeal to Lemma \ref{lemma:hadamard}.
Otherwise, fix $a \in \GG$ with order greater than $2$.
The character corresponding to $a$ is $\chi_{a}(x) = \prod_{j=1}^{k} \chi_{a_{j}}(x_{j})$. 
From Equation (\ref{eqn:group_lambda}),
\begin{equation}
\lambda_{a} 
	= \sum_{x \neq 0} f(x)\overline{\chi}_{a}(x)
	= \sum_{x \neq 0} f(x)\overline{\chi}_{-a}(-x)
	= \sum_{x \neq 0} f(-x)\overline{\chi}_{-a}(-x)
	= \lambda_{-a}.
\end{equation}
Thus, the spectral gap of $M_{\GG}^{f}$ is zero.
Finally, since $\GG$ is Abelian, its characters are complex roots of unity;
thus, applying Equation (\ref{eqn:aakv}), we obtain the claim.
\qed\\

\par\noindent The above theorem implies that the $n$-cube and the standard circulant graphs
are not average uniform mixing, as stated in the following corollary. 

\begin{corollary}
No $\Int_{2}^{n}$-circulant and no $\Int_{n}$-circulant, except for $C_{2}$, is average uniform mixing.
\end{corollary}

Next, we relax our requirement of exact uniform average mixing and allow mixing to be $(1/n)$-uniform.
We observe that the cycle graphs and the complete graphs behave differently with respect to average 
near uniform mixing.

\begin{theorem}
The cycle $C_{n}$ is average $(1/n)$-uniform mixing.
\end{theorem}
\prf
Let $\omega = \exp(2\pi i/n)$. Using Equation (\ref{eqn:aakv}) for circulants, we have
\begin{equation}
\overline{P}(\ell) 
	= \frac{1}{n^2}\sum_{j,k=0}^{n-1} \omega^{(j-k)\ell}\iverson{\lambda_{j} = \lambda_{k}}
	= \frac{1}{n} + \frac{1}{n^{2}}\sum_{j \neq k} \omega^{(j-k)\ell}\iverson{\lambda_{j} = \lambda_{k}}.
\end{equation}
A result of Diaconis and Shahshahani (see \cite{diaconis}) states that
$||\overline{P} - U|| \le \frac{1}{4}\sum_{\rho} |\widehat{\overline{P}}(\rho)|^{2}$,
where the sum is over non-trivial irreducible representations.
The characters of $\Int_{n}$ are given by $\chi_{a}(x) = \omega^{ax}$, and thus, for $a \neq 0$,
\begin{equation}
\widehat{\overline{P}}(a) = \sum_{\ell} \overline{P}(\ell)\chi_{a}(\ell) 
	= \frac{1}{n^{2}}\sum_{j \neq k: \ \lambda_{j} = \lambda_{k}} \sum_{\ell} \omega^{(j-k+a)\ell} 
	= \frac{1}{n}.
\end{equation}
The last equality holds because there is a unique pair $(j,k)$ such that $j-k+a=0$; this pair contributes $n$ to the sum while
the other pairs contribute $0$ to the sum. Therefore, 
$||\overline{P} - U|| \le (n-1)/4n^{2} < 1/4n$.
\qed\\

In contrast, the average distribution of a quantum walk on the complete graphs $K_{n}$ is not near uniform.

\begin{theorem} (Ahmadi \etal \cite{abtw03})
The complete graph $K_{n}$ is not average $(1/n)$-uniform mixing.
\end{theorem}
\prf
As shown in \cite{abtw03}, for $\ell \neq 0$, we have $\overline{P}(\ell) = 2/n^{2}$, 
and $\overline{P}(0) = 1-2(n-1)/n^{2}$. Thus $||\overline{P}-U|| = 2(1-1/n)(1-2/n) \gg 1/n$.
\qed


\section{Conclusions}

In this work, we have shown that a continuous-time quantum walk on cycles exhibit strong non-classical mixing
characteristics. First, we prove that a continuous-time quantum walk on most even-length cycles is not instantaneous
uniform mixing. This partially settles a conjecture made in \cite{abtw03}. Second, we prove that a continuous-time 
quantum walk on any cycle is not average uniform mixing. The latter result is obtained as a corollary of a stronger 
theorem for a continuous-time quantum walk on any Abelian circulant graph. This class of graphs include the natural 
families of cycles, complete graphs, hypercubes, and others. In contrast, classical lazy random walks on the same 
graphs mix to the uniform distribution.

We leave the case of the odd-length cycles as well as cycles of length $2^{u}q$, with $u \ge 1$ and $q \equiv 1\pmod{4}$,
for future work. Also, we are curious to investigate if there is something interesting about the quantum mixing status 
of prime-length cycles. Finally, the only known class of graphs with instantaneous uniform mixing is the hypercube family, 
and the only graph known to be average uniform mixing is the connected two-vertex graph. It is conceivable that these are 
{\em the} two lone examples of uniform mixing in circulants.


\section*{Acknowledgments}

This research was supported in part by the National Science Foundation grants 
DMS-0097113, DMS-0646847, DMR-0121146 
and also by the
National Security Agency grant 42642,
while the authors were part of the Clarkson-Potsdam Research Experience for Undergraduates 
(REU) Summer program in Mathematics 2003 and 2007.


\end{document}